# Recommendations from the Ad Hoc Committee on SETI Nomenclature

Jason T. Wright[1], Sofia Sheikh[1], Iván Almár[2], Kathryn Denning[3], Steven Dick[4] and Jill Tarter[5]

The Ad Hoc Committee on SETI Nomenclature was convened at the suggestion of Frank Drake after the Decoding Alien Intelligence Workshop at the SETI Institute in March 2018. The purpose of the committee was to recommend standardized definitions for terms, especially those that are used inconsistently in the literature and the scientific community. The committee sought to recommend definitions and terms that are a compromise among several desirable but occasionally inconsistent properties for such terms:

1. Consistency with the historical literature and common use in the field
2. Consistency with the present literature and common use in the field
3. Precision of meaning
4. Consistency with the natural (i.e. everyday, non-jargon) meanings of terms
5. Compatibility with non-English terms and definitions

The definitions below are restricted to technical, SETI contexts, where they may have jargon senses different from their everyday senses. In many cases we include terms only to deprecate them (in the sense of "to withdraw official support for or discourage the use of…in favor of a newer or better alternative", Merriam-Webster sense 4).

This is a consensus document that the committee members all endorse; however, in many cases the individual members have (or have expressed in the past) more nuanced opinions on these terms that are not fully reflected here, for instance Almár (2008, Acta Astronautica, **68**, 351), Denning (2008, NASA-SP-2009-4802 Ch. 3 pp.63–124), and Wright (2018, arXiv:1803.06972).

---

[1] Penn State University Center for Exoplanets and Habitable Worlds and Department of Astronomy & Astrophysics, PA, USA
[2] Hungarian Academy of Sciences, Budapest, Hungary
[3] Associate Professor, York University Department of Anthropology, Toronto, ON, Canada
[4] Former NASA Chief Historian and Former Baruch S. Blumberg NASA/Library of Congress Chair in Astrobiology
[5] Chair Emerita for SETI Research, SETI Institute, Mountain View, CA, USA

**SETI**

**1.** *n.* A subfield of astrobiology focused on searching for signs of non-human technology or technological life beyond Earth. The theory and practice of searching for extraterrestrial technology or technosignatures.
**2.** *adj.* Of that subfield
**3.** *acronym* (the) Search for Extraterrestrial Intelligence

*Notes: SETI* is an acronym and so should be pronounced (sɛ'ti in American English), not spelled out.

Especially at NASA, *SETI* has sometimes been more narrowly defined to refer specifically to radio SETI, or to a particular program at NASA. The term generally has broader scope in the SETI literature and community.

*SETI* should not be used as shorthand for the SETI Institute, which is an independent entity and should be referred to by its full name to avoid confusion.

---

**Extraterrestrial**

When referring to that which SETI seeks, shorthand for life or technology not originating recently on Earth.

*Notes:* The terminology is complicated by the possibility that Earth life originated elsewhere or has significantly spread beyond our planet (i.e. panspermia hypothesis), and the fact that human technology is present throughout our Solar System.

Although the term in general has broader meaning than this, even in astronomy, its use in the SETI context to refer to non-terrestrial life is so entrenched that we endorse this jargon sense and will not attempt to refine it to be more precise. Alternatives, such as *non-terrestrial,* have similar difficulties and are not yet in wide use in the SETI community.

So by this definition, life on another planet with a common origin to Earth life but which diverged billions of years ago would be *extraterrestrial*, but Earth life accidentally brought to Mars on a human-built lander would not. Similarly, humanity's solar system probes are not *extraterrestrial technology* by this jargon definition.

---

**Intelligence**

In the acronyms SETI and ETI, the quality of being able to deliberately engineer technology which might be detectable using astronomical observation techniques.

*Notes*: Definitions of intelligence are slippery and much broader than *technological* and so we recommend against the term's use as a synonym for *technological* in general. Its use in the acronym SETI is sufficiently entrenched that we recommend against a more precise rebranding of the field.

---

**Extraterrestrial Technological Entities** or
**Extraterrestrial Technological Species** or
**Extraterrestrial Intelligence** or
**ETI**

The entities that created the technosignatures SETI seeks to find.

*Notes*: To be preferred to *alien race* or *alien civilization* because of human connotations in those terms. *ETI* is not ideal because it contains *intelligence* but is acceptable because it directly references the term *SETI*.

We note that some rarely seen alternatives such as *sophont* (literally "wise one") and *technologist* (those responsible for technosignatures) are useful because they lack connotations about the entities' nature except their ability to create technology. However, none has yet gained sufficient traction in the community to earn our endorsement. *Extraterrestrial society* may be preferable to *species* because it does not assume a narrow biological species definition, but is not in wide use in the literature.

---

**Alien**

**1. n.** Deprecated term for extraterrestrial species
**2. adj.** Term to be used with care, meaning of or pertaining to extraterrestrial species or their technology

*Notes*: As a noun, to be avoided for many reasons, including its associations with unscientific portrayals of extraterrestrial life in popular culture and its legal meaning in relation to immigration.

We do not recommend the use of *alien* as an adjective in general, but acknowledge it is a common and useful term, so instead recommend that the term be used with care and consideration of the issues we note with the noun. That is, *alien species* is to be preferred to *alien*, but *extraterrestrial species* is better.

---

**Civilization**

In a SETI context, e.g. (*extraterrestrial* or *alien civilization*) usually synonymous with *technological species*. Use with care.

*Notes:* The term *civilization* has imprecise popular meanings, but also particular scholarly meanings in relation to human history that are not generally what is meant by the term in a SETI context. Because of its ambiguity and anthropocentrism, the term is a suboptimal synonym for *technological species,* but it is nonetheless widely used in the literature. *Society* is a good alternative but not yet in common use.

---

**Advanced** (technology, species, etc.)

Deprecated term describing a technology or species that manipulates energy or matter in a manner or extent which surpasses humanity's capabilities.

*Notes*: This term unhelpfully echoes deprecated theories of human history which rank human societies from "primitive" to "advanced" based on ill-defined and ethnocentric measures. While acknowledging the cumulative nature of human science and technology, we recommend instead simply specifying the scale or nature of the technology referenced (for instance, using the Kardashev scale, physical size, etc., as context requires).

**Technology**

The physical manifestations of deliberate engineering. That which produces a technosignature.

**Technosignature**

Any sign of technology that is not also a biosignature (i.e. that is not also a sign or byproduct of metabolism in common with non-technological species).

*Notes*: By analogy with *biosignature*, *technosignature* refers to signs or signals, which, if observed, would allow us to infer the existence of technological life elsewhere in the universe; this tracks the natural meaning of the word and the intent of its original coinage (see Tarter (2007) [Highlights of Astronomy **14**, 14](#)). In some cases, the line between biosignatures and technosignatures may be unclear, and technology need not be of biological origin, as in the case of "postbiological evolution."

---

**Natural**

Not being a product of deliberate engineering; antonym of *artificial*.

**Artificial**

Being a product of deliberate engineering; antonym of *natural*.

*Notes*: These terms are important because they are part of the boundaries between SETI and the rest of astrobiology. Unfortunately, they are slippery, and are not even well defined for observable phenomena on Earth. For instance, a beaver's dam or the tunnels under an anthill are somewhat ambiguous by these definitions. It is also unclear to what degree we will recognize extraterrestrial artifice as such if and when we find it.

---

**Beacon**

Any "we are here" sign or signal deliberately engineered by a technological species to be noticed, recognized, and understood by another technological species as evidence or proof of the first technological species' presence.

*Notes*: This is useful because searches for such signals are amenable to game theoretical approaches (i.e. magic frequencies, Schelling points, etc.). It also matches the natural meaning of the term. Sometimes the term is used to refer only to content-free signals (as in the original meaning of "beacon", i.e. a lighthouse or signal fire), but we recommend other terms for this sense (see **dial tone**).

---

**Dial Tone** or
**Door bell**

A content-free beacon, i.e. one that communicates *only* the existence of technological life.

*Notes*: This concept may be useful in discussions of the social consequences of detection, because human reactions to an extraterrestrial signal without content may be quite different from reactions to the contents of an actual extraterrestrial message. These terms for the concept are perhaps too colloquial and parochial (in particular *dial tone* may become obscure in the near future) but we include them because they convey the sense of the concept well by analogy to their common meanings, and because they are already in use in the SETI community.

---

**Settle** or
**Colonize** (and their cognates)

The (hypothetical) spread of a species across space, in particular throughout a planetary system, among stars, and throughout a galaxy.

Notes: *Colonize* is the more common term in the SETI literature, but some avoid (or embrace) its use because of its connotations of the global colonial exploits of European powers.

**Artifact SETI**

**1. n.** A subfield of SETI dedicated to the search for physical manifestations of technology, exclusive of communicative transmissions.

*Notes*: This sense is useful because it captures many approaches to SETI which differ in strategy or philosophy from searches for deliberate transmissions meant for communicative purposes. Note that searches for communicative transmissions are often also sensitive to non-communicative transmissions; for instance, radio searches may be sensitive to extraterrestrial radar signals or signs of propulsion.

*Artifact SETI* includes searches for probes in the Solar System (which are sometimes called Solar System SETI or Probe SETI), and searches beyond our Solar System for atmospheric technosignatures, signs of propulsion, and waste heat from industry.

*Artifact SETI* often appears in the SETI literature in a narrower sense, to refer to searches for artifacts in the Solar System, exclusively. We recommend the broader sense because it is useful, closer to the natural meaning of the term, and already in use in the SETI literature.

In the literature, *artifact SETI* is sometimes called *Dysonian SETI*, but that term more naturally refers to searches for specific technosignatures suggested by Freeman Dyson, especially Dyson Spheres.

There is no common term in the literature for searches for communicative transmissions, specifically. *Communication SETI* appears in the literature but is easily confused with CETI. Both it and *transmission SETI* have ambiguous natural meanings and are easily confused with METI.

---

**SETA**

**1. n.** Deprecated term for the search for physical artifacts, such as probes in the Solar System
**2. acronym** (the) Search for Extraterrestrial Artifacts

*Notes*: This term was originally used to distinguish searches for Solar System artifacts from searches for communicative transmissions. Because we endorse a broader definition of SETI, we recommend that such searches be considered a subset of SETI, not a distinct activity, and so we deprecate SETA, which is not in wide use today, anyway.

---

**METI**

**1. n.** The activity of humans eliciting contact with technological species via the transmission from Earth of beacons or other invitations for contact (including the Arecibo messages and Pioneer plaques).
**2. acronym** Messaging Extraterrestrial Intelligence

*Notes:* The word is an acronym and so should be pronounced (mɛ'ti in American English), not spelled out.

METI is a controversial activity. Some consider it to be logically continuous with SETI, and others consider it to be a distinct activity. To some it also includes replies to future hypothetical incoming transmissions, and theoretical work on how to communicate with ETI, but others consider these to be distinct from METI.

*METI* should not be used as shorthand for METI International, which is an independent entity and should be referred to by its full name to avoid confusion.

---

**Active SETI**

Deprecated term for *METI*

*Notes*: Avoid this term because of syntactic ambiguity (e.g. "Dr. X is PI of three active SETI programs", "Dr. Y is an active SETI researcher")

**Schelling Point**

(game theory) An equilibrium in a non-communicative cooperative game such as a mutual search; i.e. a mutually obvious game strategy. The most obvious example is the consideration of "magic frequencies" for radio beacons, such as the 21 cm line in radio astronomy.

*Notes*: The identification of magic frequencies in SETI as a quintessential example of this important game theoretical concept was established by Thomas Schelling himself in 1960. The term is synonymous with the jargon term *focal point* in game theory, but *Schelling point* is to be preferred in astronomical contexts to avoid confusion with the term in optics.

*Schelling point* has priority over and is to be preferred to terms in the literature that have not caught on such as *mutual strategy of search*, *synchrosignals*, or *convergent strategy of mutual search*.

---

**CETI**

**1. n.** Deprecated term encompassing: 1) SETI, 2) the theory of encoding and decoding interstellar messages, and 3) METI.
**2. acronym** Communication with Extraterrestrial Intelligence

*Notes:* The term is rarely seen any more, and so we deprecate it in favor of *SETI*.

---

**Kardashev scale**

The Kardashev Scale measures the energy supply of an entire technological society in one of three integer types, as described by Kardashev in 1964 in *Soviet Astronomy* **8**, 217. Other senses of the term, including extensions of the scale, should be explicitly described or cited.

---

**Fermi Paradox**

There is no canonical formulation of the Fermi Paradox (see Gray (2015) [_Astrobiology_ 15, 195](#)) but its earliest invocations refer to the (supposed) inconsistency between hypothetical timescales for settlement of the Galaxy and the absence of extraterrestrial spacecraft or probes on Earth. When used, the intended meaning should be given explicitly.

---

**Drake Equation**

Proper citation to most common form of the equation, which describes the number of potentially detectable sources of radio transmissions in the Galaxy, is
Drake, F (1965) The Radio Search for Intelligent Extraterrestrial Life, in [_Current Aspects of Exobiology_, ed. G Mamikunian and M. H. Briggs](#), 323-45. Oxford University Press.

When using the term in other contexts, define precisely what $N$ is intended to represent.

---

**Rio Scale**

A scale from 1–10 developed by astronomers to express their estimates of the importance of a report of detection or contact with an extraterrestrial species, computed as the product of a claim's credibility and level of consequences. The scale was adopted by the International Academy of Astronautics SETI Permanent Committee and is currently at version 1.2, available at http://www.setileague.org/iaaseti/riocalc.htm.

_Notes:_ An updated scale, "Rio 2.0" was recently proposed by Forgan et al. ([_International Journal of Astrobiology, accepted_](#)) but has not been adopted by the IAA and so currently has no official status.